\documentclass[aps,twocolumn,showpacs,preprintnumbers,amsmath,amssymb,superscriptaddress]{revtex4}

\usepackage{graphicx}
\usepackage{dcolumn}
\usepackage{bm}
\usepackage{epsfig}

\begin{document}

\preprint{APS/123-QED}

\title{$^{110,116}$Cd($\alpha,\alpha$)$^{110,116}$Cd elastic scattering and systematic investigation of elastic $\alpha$ scattering cross sections
  along the $Z$ = 48 isotopic and $N$ = 62 isotonic chains}

\author{G. G. Kiss}%
 \email{ggkiss@atomki.hu}
 \altaffiliation [present address:]{ Laboratori Nazionali del Sud, INFN, Catania, Italia}
\affiliation{%
Institute of Nuclear Research (ATOMKI), H-4001 Debrecen, POB.51., Hungary}%
\author{P. Mohr}%
\affiliation{%
Institute of Nuclear Research (ATOMKI), H-4001 Debrecen, POB.51., Hungary}%
\affiliation{%
Diakonie-Klinikum, D-74523 Schw\"abisch Hall, Germany}%
\author{Zs.\,F\"ul\"op}%
\affiliation{%
Institute of Nuclear Research (ATOMKI), H-4001 Debrecen, POB.51., Hungary}%
\author{Gy.\,Gy\"urky}%
\affiliation{%
Institute of Nuclear Research (ATOMKI), H-4001 Debrecen, POB.51., Hungary}%
\author{Z. Elekes}%
\affiliation{%
Institute of Nuclear Research (ATOMKI), H-4001 Debrecen, POB.51., Hungary}%
\author{J. Farkas}%
\affiliation{%
Institute of Nuclear Research (ATOMKI), H-4001 Debrecen, POB.51., Hungary}%
\author{E. Somorjai}%
\affiliation{%
Institute of Nuclear Research (ATOMKI), H-4001 Debrecen, POB.51., Hungary}%
\author{C. Yalcin}%
\affiliation{%
Institute of Nuclear Research (ATOMKI), H-4001 Debrecen, POB.51., Hungary}%
\affiliation{%
Kocaeli University, Department of Physics, TR-41380 Umuttepe, Kocaeli, Turkey} %

\author{D. Galaviz}%
\altaffiliation [present address:]{ Centro de F\'isica Nuclear da Universidade de Lisboa, 1649-003, Lisbon, Portugal}
\affiliation{%
Instituto de Estructura de la Materia, CSIC, E-28006 Madrid, Spain}%

\author{R. T.  G\"uray}%
\affiliation{%
Kocaeli University, Department of Physics, TR-41380 Umuttepe, Kocaeli, Turkey}%
\author{N. \"Ozkan}%
\affiliation{%
Kocaeli University, Department of Physics, TR-41380 Umuttepe, Kocaeli, Turkey}%

\author{J. G\"orres}%
\affiliation{%
University of Notre Dame, Notre Dame, Indiana 46556, USA}%

\date{\today}

\begin{abstract}
The elastic scattering cross sections for the reactions
$^{110,116}$Cd($\alpha,\alpha$)$^{110,116}$Cd at energies above and below  the
Coulomb barrier are presented to provide a sensitive test for the
alpha-nucleus optical potential parameter sets. Additional constraints for the
optical potential are taken from the analysis of elastic scattering excitation
functions at backward angles which are available in literature.
Moreover, the variation of the elastic alpha scattering cross sections along
the $Z = 48$ isotopic and $N = 62$ isotonic chain is investigated by the study
of the ratios of the  of
$^{106,110,116}$Cd($\alpha,\alpha$)$^{106,110,116}$Cd scattering cross sections at E$_{c.m.}
\approx$ 15.6 and 18.8 MeV and the ratio of the
$^{110}$Cd($\alpha,\alpha$)$^{110}$Cd and
$^{112}$Sn($\alpha,\alpha$)$^{112}$Sn reaction cross sections at E$_{c.m.}
\approx$ 18.8 MeV, respectively. These ratios are sensitive probes for the
alpha-nucleus optical potential parameterizations. 
The potentials under study are a basic
prerequisite for the prediction of $\alpha$-induced reaction cross sections,
e.g.\ for the calculation of stellar reaction rates in the astrophysical $p$-
or $\gamma$-process.

\end{abstract}


\maketitle

\section{Introduction}
\label{sec:intro}
Most of the nuclei heavier than iron are built up via neutron capture reactions in the
so-called $s$ and $r$ processes. However, on the proton-rich side of the
valley of stability there are 35 proton-rich nuclei not created by neutron capture processes. 
These mostly even-even
proton-rich, stable isotopes between $^{74}$Se and $^{196}$Hg are the
so called $p$ nuclei \cite{woo78}. Their natural isotopic abundance is 10 -
100 times less than that of the more abundant neutron-rich isotopes which were
synthesized in the $s$- or $r$-processes.

\begin{center}
\begin{table*}
\caption{\label{tab:iso} 
  Charge and neutron number, energy of the first excited
  state of the target nuclei, enrichment, and E$_{lab.}$ and E$_{c.m.}$
  energies for each of the angular distributions studied in the present
  work, the $^{106}$Cd and $^{112}$Sn data are taken from \cite{gal_phd, kis06,gal05}. The nuclear data are from \cite{nndc106, nndc110, nndc116, nndc112}.} 
\setlength{\extrarowheight}{0.2cm}
\begin{tabular}{cccccccc}
\hline
\parbox[t]{1.3cm}{\centering{target \\ nuclei }} &
\parbox[t]{1.3cm}{\centering{proton \\ number }} &
\parbox[t]{1.3cm}{\centering{neutron \\ number}} &
\parbox[t]{2.3cm}{\centering{1st excited \\ state [keV]}} &
\parbox[t]{2.3cm}{\centering{enrichment [\%] }} &
\parbox[t]{2.3cm}{\centering{E$_{lab.}$ [MeV] }} &
\parbox[t]{2.3cm}{\centering{E$_{c.m.}$ [MeV] }} &
\parbox[t]{1.0cm}{\centering{Ref. }} \\
\hline
 $^{110}$Cd & 48 & 62  & 657.76 & 95.7 &  16.14, 19.46 & \centering{15.6, 18.8}    &  this work\\
 $^{116}$Cd & 48 & 68  & 513.49 & 98.3 &  16.14, 19.46 & \centering{15.6, 18.8}    &  this work\\ 
\hline 
$^{106}$Cd & 48 & 58  & 632.64 & 96.5 & 16.13, 19.61 &  \centering{15.6, 18.9}    & \cite{gal_phd, kis06} \\
 $^{112}$Sn & 50 & 62  & 1256.85 & 99.8 & 19.51 & \centering{ 18.8}              &  \cite{gal05}\\ 
\hline
\end{tabular}
\end{table*}
\end{center}

In the production of the $p$ nuclei, photon-induced reactions at temperatures
around a few GK are playing a crucial role. It is generally accepted that the
main stellar mechanism synthesizing these nuclei --- the so called $\gamma$ process
--- is initiated by ($\gamma$,n) photodisintegration reactions on preexisting
neutron-rich $s$ and $r$ seed nuclei. Photons with high energy and high flux
--- necessary for the $\gamma$ - induced reactions --- are available in explosive
nucleosynthesis scenarios like in the Ne/O burning layer in type II supernovae
\cite{woo78, arn03}. As the neutron separation energy increases along the
($\gamma$,n) path toward more neutron deficient isotopes, ($\gamma$,p) and
($\gamma$,$\alpha$) reactions become more important and process the material
toward lower atomic numbers \cite{arn03,rau06,rap06, rau10}. Recently, consistent studies of
$p$ nucleosynthesis have become available, employing theoretical reaction
rates in large reaction networks \cite{rau06,rap06}. These studies confirmed
that, in the case of the heavy $p$ nuclei (140 $\leq A \leq$ 200),
($\gamma$,n) and ($\gamma,\alpha$) reactions play the dominant role.

Modeling the synthesis of the \textsl{p} nuclei and calculating their
abundances requires an extended reaction network involving more than 10$^4$
reactions on 2000 mostly unstable nuclei. The reaction rates are usually based
on the Hauser-Feshbach statistical model. Because of the experimental
challenges very few ($\gamma,\alpha$) studies have been performed until now: in
a pioneer experiment the cross section of the
$^{144}$Sm($\gamma,\alpha$)$^{140}$Nd has been measured recently
\cite{nai08}. However, in such an experiment the target nucleus is always in
its ground state, whereas in stellar environments thermally populated excited
states also contribute to the reaction rate. Thus theoretical considerations
cannot be avoided \cite{moh08}. Alternatively, the ($\gamma,\alpha$) rates can
be determined experimentally by measuring the inverse ($\alpha$,$\gamma$)
reaction cross section and converting the results by using the detailed
balance theorem. In this direction the influence of thermally excited states
remains relatively small \cite{moh08,rau09}.
In recent years a range of ($\alpha,\gamma$) reaction cross
sections on $^{70}$Ge, $^{96}$Ru, $^{106}$Cd, $^{112,117}$Sn, $^{113}$In,
$^{144}$Sm, $^{151}$Eu and $^{169}$Tm has been measured using the activation
method, and the results have been compared with model predictions
\cite{ful96,rap01,gyu06,ozk07,cat08,yal09,som98, gyu10, kis10}.

It was generally found that the ($\gamma,\alpha$) and ($\alpha,\gamma$)
reaction cross section calculations are very sensitive to the choice of the
$\alpha$-nucleus potential which is a sum of a Coulomb and a nuclear part
(the latter one consists of a real and an imaginary part). The cross section
predictions using different global alpha-nucleus optical potential
parameterizations can differ within an order of magnitude \cite{ozk07}. Since
the parameters of the global alpha-nucleus optical potentials are usually
determined from the analysis of the angular distributions of elastically
scattered alpha particles (and are adjusted to alpha-induced cross sections if
experimental data exist), the elastic alpha scattering cross sections on
several $p$ nuclei had been measured in recent years at ATOMKI
\cite{moh97,ful01,gal_phd,kis06,gal05} and similar experiments are ongoing at Notre Dame University \cite{pal08}.

In order to increase our knowledge on the alpha-nucleus optical potential
parameterizations the energy and the mass dependence of the potential
parameters has to be understood. Although it would be helpful to perform
systematic investigations on the alpha-nucleus optical potential
parameterizations in the whole mass range of the $p$ nuclei (i.e.\ from about $A \approx 70$ up to almost $A \approx 200$), the fact that
most of these nuclei have low-lying first excited state makes a study of elastic scattering
experimentally very difficult. Experimental studies are well accessible in the
region of the lower mass $p$ nuclei in the $A \approx 100$ mass range and
around $A \approx 140 - 150$ where relatively high-lying first excited states
are found. Here the features of the optical potential
parameterizations should be as well understood as possible. However, as a word
of caution, one should to keep in mind that high-lying first
excited states are related to shell closures (e.g.\ $Z=50$ and $N=82$ in these
mass regions), and the imaginary part of the optical potential is typically
smaller for closed-shell nuclei than for nuclei off closed shell. It is
one motivation of the present investigation to study the optical potential for
nuclei without closed shells.

From the astrophysical point of view, the potential parameters should be
derived in the relevant energy region, in the so called Gamow window. However,
at those sub Coulomb energies the elastic scattering cross sections are practically not
deviating from the Rutherford cross section, and for this reason it is not
possible to derive reliable optical potential parameters for these energies. Consequently, the
experiments have to be performed at slightly higher energies just below and above the Coulomb barrier and then the
resulting optical potential parameters have to be extrapolated down to the
relevant energy region. Contrary to the real part of the nuclear potential
which has a smooth energy-dependence, the imaginary part changes drastically
around the Coulomb barrier. 

A global alpha-nucleus optical potential must be able not only to
provide a correct prediction for the alpha elastic scattering angular
distributions but also to describe the variation of the angular distributions
along isotopic and isotonic chains.  This is especially important for the
extrapolation to unstable nuclei where no measured alpha-induced reaction data
are available and the potential cannot be derived from experimental scattering
data. Recently, the variation of the scattering cross sections along the $Z =
50$ isotopic and $N = 50$ isotonic chain has been investigated. The
ratio of the measured cross sections of the
$^{112}$Sn($\alpha,\alpha$)$^{112}$Sn/$^{124}$Sn($\alpha,\alpha$)$^{124}$Sn
($Z = 50$) and
$^{89}$Y($\alpha,\alpha$)$^{89}$Y/$^{92}$Mo($\alpha,\alpha$)$^{92}$Mo ($N
=50$) reactions showed an oscillation pattern at backward angles. It was found
that both regional and global alpha-nucleus optical potential parameterizations
failed to reproduce these oscillation patterns \cite{gal05,kis09}.

In order to further investigate the variation of the elastic alpha scattering
cross sections along isotopic and isotonic chains, in the present work the
$^{110,116}$Cd($\alpha,\alpha$)$^{110,116}$Cd reactions are studied at
energies above and below the Coulomb barrier. This paper is organized as
follows. In Sec. \ref{sec:exp} we describe our experimental procedure. The
measured angular distributions are compared to predictions using local
(Sec.~\ref{sec:local}), regional and global optical potential
parameterizations in Sec.~\ref{sec:global}. Excitation functions taken from
literature \cite{eis74,bad78,mil81} provide further information on the
potentials; the experimental excitation functions are compared to the
results from the local, regional, and global potentials in
Sec.~\ref{sec:exci}. Additionally, all calculations are used to predict the
ratio of angular distributions along the cadmium ($Z = 48$) 
isotopic and $N = 62$ isotonic chains (Sec.~\ref{sec:ratio}). The elastic
alpha scattering cross sections of the $^{106}$Cd($\alpha,\alpha$)$^{106}$Cd
and the $^{112}$Sn($\alpha,\alpha$)$^{112}$Sn are taken from
\cite{gal_phd,kis06,gal05}. A further detailed study on
$^{106}$Cd($\alpha,\alpha$)$^{106}$Cd elastic scattering and the influence of
the chosen potential on $\alpha$-induced cross sections of $^{106}$Cd will be
presented in a separate paper \cite{gal11}.

\section{Experimental procedure}
\label{sec:exp}
The experiment was carried out at the cyclotron laboratory of ATOMKI,
Debrecen. A similar experimental setup was used also in the previous
experiments \cite{moh97, ful01, kis06, gal05, kis09} and is described in more
detail in \cite{mat89, kis08}. The proton and neutron number, the energy of
the first excited states of the target nuclei and the energies of the measured
angular distributions are summarized in Table \ref{tab:iso}. The following
paragraphs provide a short description of the experimental setup.

\begin{center}
\begin{figure}
\resizebox{0.86\columnwidth}{!}{\rotatebox{0}{\includegraphics[clip=]{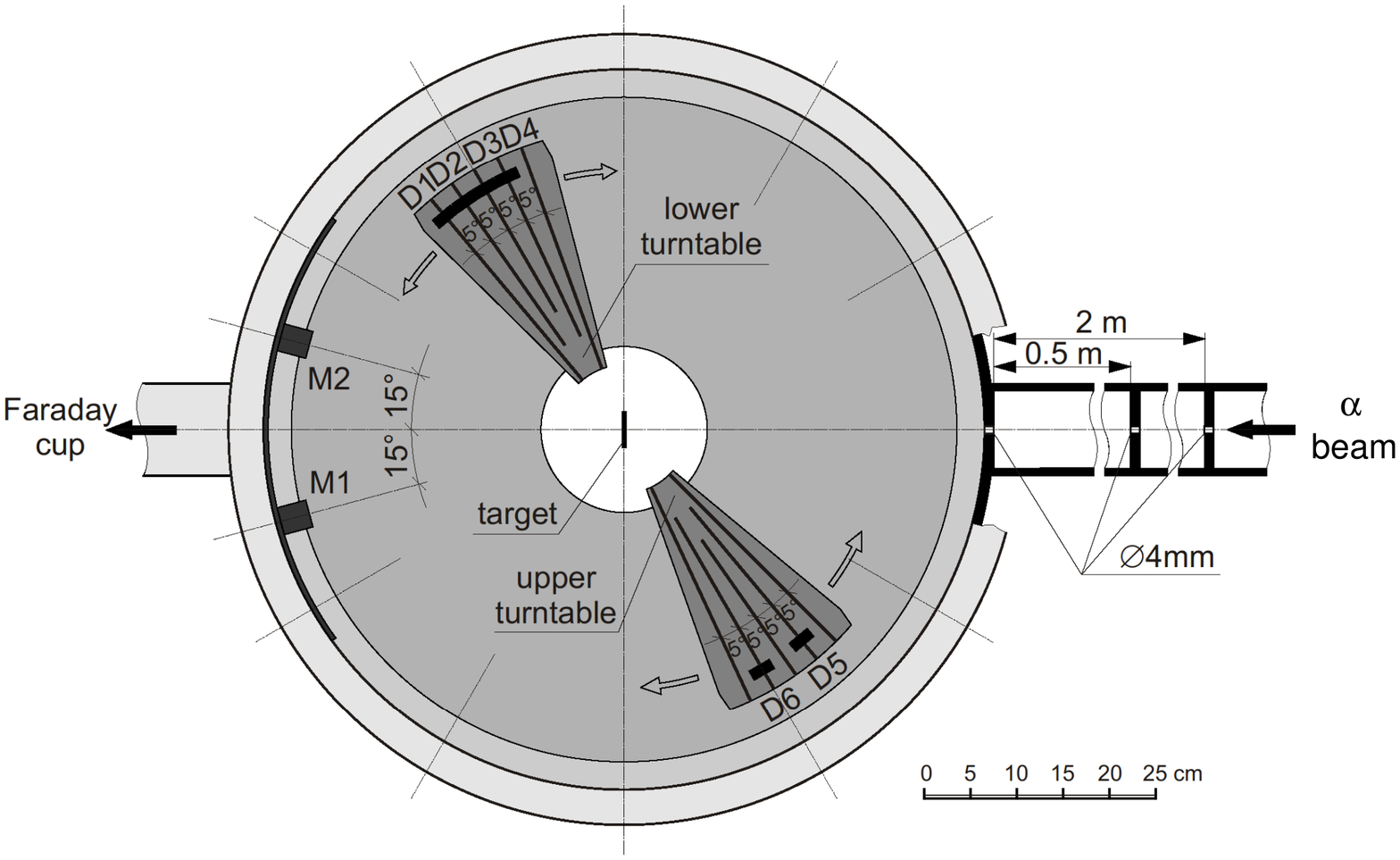}}}
\caption{\label{fig:chamber} Schematic view of the scattering chamber with
  monitor detectors (M1 and M2, mounted in the wall of the chamber at fixed
  $\pm$ 15$^{\circ}$ angles) and detectors (D1-D6 on the two turntables) used to
  measure the yield of the elastically scattered alpha particles. For details,
  see text.}
\end{figure}
\end{center}

The targets were produced by evaporating highly enriched ($\geq 95\%$)
$^{110,116}$Cd onto thin carbon foils ($\approx$ 20 $\mu$g/cm$^{2}$). The
target thickness was approximately 200 $\mu$g/cm$^{2}$, determined via alpha particle energy loss measurement using radioactive sources. The targets were
mounted on a remotely controlled target ladder in the center of the scattering
chamber. Figure \ref{fig:chamber} illustrates the scattering chamber.

\begin{center}
\begin{figure}
\resizebox{0.86\columnwidth}{!}{\rotatebox{0}{\includegraphics[clip=]{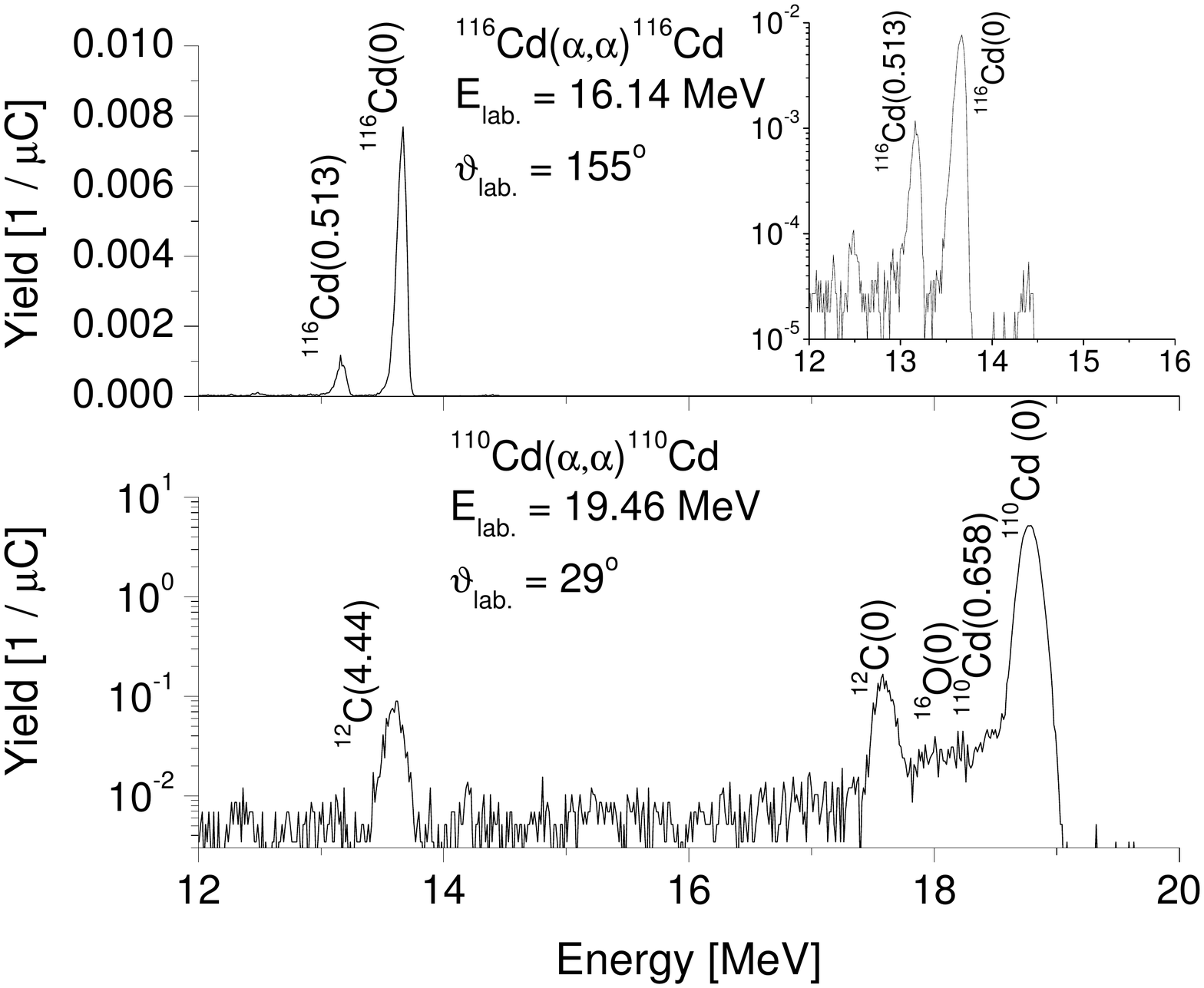}}}
\caption{\label{fig:spec} Typical spectra of the
  $^{116}$Cd($\alpha,\alpha$)$^{116}$Cd and
  $^{110}$Cd($\alpha,\alpha$)$^{110}$Cd reactions at E$_{lab.}$ = 16.14 MeV
  (upper part) and E$_{lab.}$ = 19.46 MeV (lower part), respectively. The upper
  spectrum was taken at $\vartheta_{lab.}$ = 155$^{\circ}$ (the inset shows the spectrum on logarithmic scale), the lower one at $\vartheta_{lab.}$ = 29$^{\circ}$.
  It can be
  seen that there are more than two orders of magnitude difference in the
  elastic scattering cross sections. The peak from elastic $^{116}$Cd-$\alpha$
  and $^{110}$Cd-$\alpha$ scattering is well resolved from the inelastic
  events and from both the $^{12}$C-$\alpha$ and $^{16}$O-$\alpha$ elastic
  scattering.}
\end{figure}
\end{center}

The energies of the alpha beam were 16.14 and 19.46 MeV with typical beam
currents of 150-200 pnA. An aperture of 2 x 6 mm was mounted on the target
ladder to check the beam position and size of the beamspot before and after
every change of the beam energy or current. We optimized the beam until not
more than 1\% of the total beam current could be measured on this aperture. As
a result of the procedure, the horizontal size of the beamspot was below 2 mm
during the whole experiment which is crucial for the precise determination of
the scattering angle. Since the imaginary part of the optical potential
depends sensitively on the energy, it is important to have a well-defined beam energy. Therefore the beam
was collimated by tight slits (1 mm wide) after the analyzing magnet; this  
corresponds to an overall energy spread of around 100 keV which is the
dominating contribution of the energy resolution of the spectra (see
Fig.~\ref{fig:spec}).

Six ion implanted silicon detectors with active areas of 50 mm$^2$ were used
to measure the angular distributions. Their solid angles varied between 1.45 x 10$^{-4}$  and 1.87 x 10$^{-4}$. The detectors were mounted on two
turntables. Two detectors with angular distance of 10$^\circ$ were mounted onto
the upper turntable and were used to measure the scattering cross sections at
forward angles. To measure the cross sections at backward angles four
detectors with angular distance of 5$^\circ$ were used. The ratio of their
solid angles was determined by measurements at overlapping angles with good
statistics ($\leq$1\% uncertainty). Typical spectra are shown in
Fig.~\ref{fig:spec}. As can be seen, the relevant peaks from elastic
$^{110,116}$Cd-$\alpha$ scattering are well separated at all angles from
elastic and inelastic peaks of target contaminations as well as from the
inelastic alpha scattering on Cd isotopes. In addition, two detectors were
mounted at a larger distance on the wall of the scattering chamber at fixed
angles $\vartheta$=$\pm$15$^\circ$ left and right to the beam axis. These
detectors were used as monitor detectors --- their solid angles were 1.1 x 10$^{-6}$  --- during the whole experiment to normalize
the measured angular distribution and to determine the precise position of the
beam on the target.

\begin{center}
\begin{figure}
\resizebox{1.0\columnwidth}{!}{\rotatebox{0}{\includegraphics[clip=]{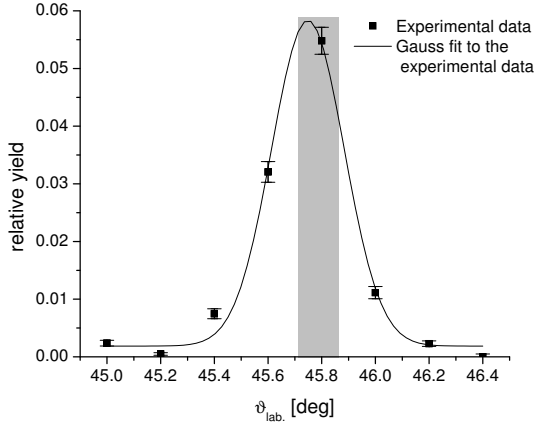}}}
\caption{\label{fig:angcalib} Relative yield of $^{12}$C recoil nuclei in
  coincidence with elastically scattered $\alpha$ particles. A Gaussian fit
 fitted  to the experimental data is shown to guide the eye. The shaded area
  represents the angular uncertainty.} 
\end{figure}
\end{center}

Knowledge of the exact angular position of the detectors is of crucial
importance for the precision of a scattering experiment since the Rutherford
cross section depends sensitively on the angle specially at forward directions. The
uncertainty in the angular distribution is dominated by the error of the
scattering angles in the forward region. To determine the scattering angle
precisely, we measured kinematic coincidences between elastically scattered
alpha particles and the corresponding $^{12}$C recoil nuclei using a pure
carbon backing as target. One detector was placed at $\vartheta$ = 70$^\circ$,
and the signals from the elastically scattered alpha particles on $^{12}$C
were selected as gates for the other detector which moved around the expected
$^{12}$C recoil angle $\vartheta$ = 45.83$^\circ$. We repeated this process
for all detector pairs. Figure~\ref{fig:angcalib}. shows the relative yield of
the $^{12}$C recoil nuclei in coincidence with elastically scattered alpha
particles as a function of the $^{12}$C recoil angle. The final angular
uncertainty was found to be $\Delta$$\vartheta$ $\leq$ 0.12$^\circ$.

Complete angular distributions between 20$^{\circ}$ and 175$^{\circ}$ were measured at energies of E$_{lab.}$ = 16.14 and 19.46 MeV in
1$^\circ$ (20$^\circ$ $\leq$ $\vartheta$ $\leq$ 100$^\circ$) and 2.5$^\circ$ (100$^\circ$ $\leq$ $\vartheta$ $\leq$ 175$^\circ$) steps. 
The statistical uncertainties varied between 0.1\% (forward angles) and 4\%
(backward angles). The count rates \textit{N($\vartheta$)} have been
normalized to the yield of the monitor detectors
\textit{N$_{Mon.}$($\vartheta$=15$^\circ$)}: 
\begin{equation}
\left(\frac{d\sigma}{d\Omega}\right)(\vartheta)\,=\left(\frac{d\sigma}{d\Omega}\right)_{Mon.}\frac{N(\vartheta)}{N_{Mon.}}\frac{\Delta\Omega_{Mon.}}{\Delta\Omega},
\end{equation}
with $\Delta$$\Omega$ being the solid angles of the detectors. Whereas the
Rutherford normalized cross sections cover only about two orders of magnitude
between the highest (forward angles at E$_{lab.}$ = 16.14 MeV) and the lowest
cross sections (backward angles at E$_{lab.}$ = 19.46 MeV), the underlying
cross sections cover more than four orders of magnitude. Over this huge range
of cross sections almost the same accuracy of about 4-5\% total uncertainty
could be achieved. This error is mainly caused by the uncertainty of the
determination of the scattering angle in the forward region and from the
statistical uncertainty in the backward region. 

\begin{center}
\begin{figure*}
\resizebox{2.1\columnwidth}{!}{\rotatebox{0}{\includegraphics[clip=]{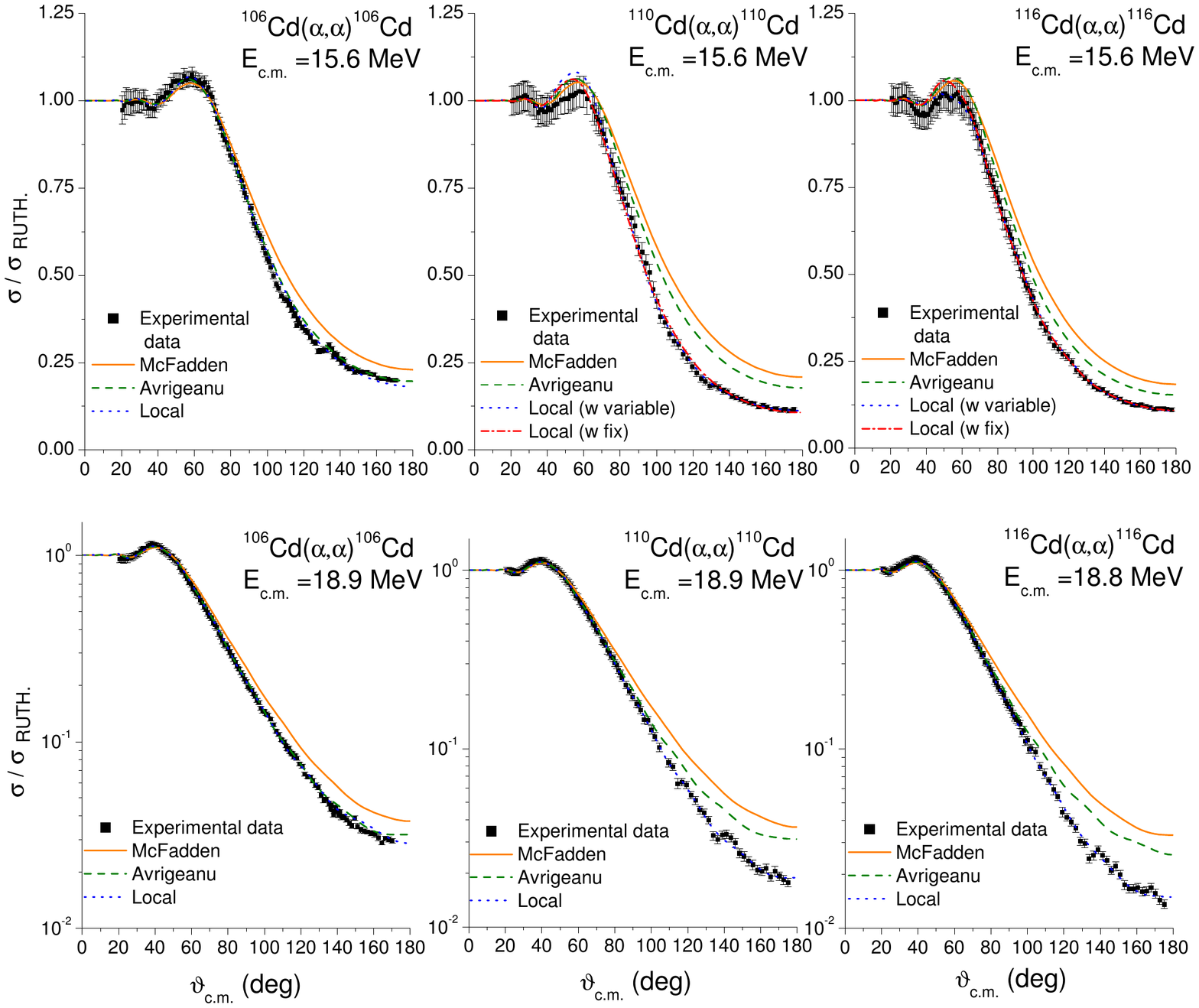}}}
\caption{
\label{fig:ruth}
(Color online.) Rutherford normalized elastic scattering cross sections of
$^{106,110,116}$Cd($\alpha,\alpha$)$^{106,110,116}$Cd reaction at E$_{c.m.}$ =
15.6 MeV ($^{106,110,116}$Cd), 18.9 MeV ($^{106}$Cd) and 18.8 MeV
($^{110,116}$Cd) versus the angle in center-of-mass frame. The lines
correspond to the predictions using the present local, regional \cite{avr09}
and global \cite{mcf} optical potential parameter sets. The
$^{106}$Cd($\alpha,\alpha$)$^{106}$Cd data were taken from \cite{gal_phd,kis06}. For
more information see Sec. \ref{sec:local} and \ref{sec:global}.} 
\end{figure*}
\end{center}

The origin of the above uncertainties has to be studied in further detail. The
uncertainty of the scattering angle is composed of two fractions. First, a
systematic uncertainty comes from the alignment of the angular scale and the
beam direction; it affects all data points in the same direction. This
uncertainty is partly compensated by the absolute normalization of the data
(see below) where the data are adjusted to Rutherford scattering at forward
angles. Second, the accuracy of setting/reading of the angle leads to a
statistical uncertainty, obviously different for each data point. The
combination of both remains below 4-5\%. From the small scatter of the data
points (see Fig.~\ref{fig:ruth}) it may be estimated that the systematic
contribution dominates the real uncertainties. Because the statistical
uncertainties are smaller than the shown error bars, it must be expected that
the resulting $\chi^2/F$ may be even below 1.0 for the locally adjusted
potentials (see Sec.~\ref{sec:local}). 

The absolute normalization is done in two steps. In a first step the absolute
normalization is taken from experiment, i.e.\ from the integrated beam
current, the solid angle of the detectors, and the thickness of the
target. This procedure has a relatively large uncertainty of the order of
$10$\,\% which is mainly based on the uncertainties of the target
thickness. In a second step a ``fine-tuning'' of the absolute normalization is
obtained by comparison to theoretical calculations at very forward angles. It
is obvious that calculated cross sections from any reasonable potential do
practically not deviate from the Rutherford cross section at the most forward
angles of this experiment; typical deviations are below 0.5\,\% for all
potentials listed in Sec. \ref{sec:local} and \ref{sec:global} (including
those potentials that do not describe details of the angular distributions at
backward angles). This ``fine-tuning'' changed the first experimental
normalization by only 2.5\,\% and thus confirmed the first normalization
within the given errors.

The measured angular distributions are shown in Fig.~\ref{fig:ruth}. The $^{106}$Cd($\alpha,\alpha$)$^{106}$Cd data, are taken from \cite{gal_phd,kis06}. The lines
are the results of optical model predictions using local, regional and global
$\alpha$-nucleus potentials (see discussion in the following
Sec. \ref{sec:local} and \ref{sec:global}). 

\section{The local $\alpha$-nucleus optical potential}
\label{sec:local}
The complex optical model potential (OMP) $U(r)$ is given by:
\begin{equation}
U(r)\,=V_{C}(r)+V(r)+iW(r),
\end{equation}

where \textit{V$_C$(r)} is the Coulomb potential, \textit{$V(r)$}, and
\textit{$W(r)$} are the real and the imaginary parts of the nuclear potential,
respectively. The volume integrals per interacting nucleon pair $J_R$ and
$J_I$ are defined as usual; although $J_R$ and $J_I$ are negative (attractive
real potential and absorption by the imaginary potential), in the discussion
the negative signs are omitted (as usual).

The V(r) real part of the local optical potential is derived from the
double-folding model. For calculating the V$_F$(r) folding potential the density-dependent M3Y interaction
\cite{kob84,sat79,abe93} was used
\begin{equation}
V(r) = \lambda \, V_{F}(r/w)
\label{eq:fold}
\end{equation}
where $\lambda \approx 1.1 - 1.4$ is the potential strength parameter
\cite{atz96} and $w = 1.0 \pm 0.05$ is the width parameter that
slightly modifies the potential width. (Larger deviations of the width
parameter $w$ from unity would indicate a failure of the folding potential.)
The nuclear densities are derived form the compilation of charge densities
measured by electron scattering \cite{vri87}. Thus, we have only two
adjustable parameters ($\lambda$ and $w$) in the real part of the potential
(e.g., compared to three parameters for Woods-Saxon potentials), and in
addition the range of these parameters is very restricted from the systematics
of volume integrals $J_R$ \cite{atz96} and the above requirement $w \approx
1$. 

The Coulomb potential is taken in the usual approximation of a homogeneously
charged sphere. The Coulomb radius $R_C$ is equal to the root-mean-square (rms)
radius of the folding potential with $w = 1$.

The imaginary part $W(r)$ of the potential is taken in the usual Woods-Saxon
parametrization. For the fits to the experimental data we use 
volume and surface potentials:
\begin{equation}
W(r) = W_V \times f(x_V) + 4 \, W_S \times \frac{df(x_S)}{dx_S}
\end{equation}
with the potential depths $W_V$ and $W_S$ of the volume and surface parts and 
\begin{equation}
f(x_i) = \frac{1}{1+\exp{(x_i)}}
\end{equation}
and $x_i = [r-R_i\,(A_T^{1/3})]/a_i$ with the radius parameters $R_i$
in the light-ion convention, the diffuseness parameters $a_i$, and $i=S,V$. It
is well established that at very low energies the surface contribution of the
imaginary part is dominating; e.g., in \cite{gal05} it is suggested that the
surface contribution is about 80\,\% for $\alpha$ scattering of the neighboring
nuclei $^{112}$Sn and $^{124}$Sn at energies below 20\,MeV. (At higher
energies, i.e.\ significantly above the Coulomb barrier, the volume
contribution is dominating.) Whereas for $^{106}$Cd a small imaginary volume
contribution was found \cite{gal_phd}, the angular distributions for
$^{110}$Cd and $^{116}$Cd can be well described using only a surface imaginary
potential. The obtained angular distributions are compared to the experimental
results and predictions from global potentials (see Sec.~\ref{sec:global}) in
Fig.~\ref{fig:ruth}, and the parameters of the fits are listed in Table
\ref{tab:local}. 
\begin{center}
\begin{table*}
\caption{
  \label{tab:local}
  Parameters of the local optical potential. The alpha optical potential
  parameters of the $^{106}$Cd and $^{112}$Sn nuclei are taken from
  \cite{gal_phd, gal05}.
} 
\setlength{\extrarowheight}{0.2cm}
\begin{tabular}{cccccccccccccccccc}
\hline
\multicolumn{1}{c}{}  
&
\multicolumn{1}{c}{}
&
\multicolumn{4}{c}{Real part}  
&
\multicolumn{6}{c}{Imaginary part}  \\

\parbox[t]{1cm}{\centering{Nucleus}} &

\parbox[t]{1cm}{\centering{E$_{c.m.}$  [MeV]}} &

\parbox[t]{0.85cm}{\centering{ $\lambda$}} &
\parbox[t]{0.85cm}{\centering{ $w$}} &
\parbox[t]{1.4cm}{\centering{ $J_R$ [MeV\,fm$^3$]}} &
\parbox[t]{0.85cm}{\centering{ $r_{R,\rm{rms}}$ [fm]}} &

\parbox[t]{0.85cm}{\centering{ W$_V$ [MeV]}} &
\parbox[t]{0.85cm}{\centering{ r$_V$ [fm]}} &
\parbox[t]{0.85cm}{\centering{ a$_V$ [fm]}} &
\parbox[t]{0.85cm}{\centering{ W$_S$ [MeV]}} &
\parbox[t]{0.85cm}{\centering{ r$_s$ [fm]}} &
\parbox[t]{0.85cm}{\centering{ a$_s$ [fm]}} &
\parbox[t]{1.4cm}{\centering{ $J_I$ [MeV\,fm$^3$]}} &
\parbox[t]{0.85cm}{\centering{ $r_{I,\rm{rms}}$ [fm]}} &

\parbox[t]{0.85cm}{\centering{ $\sigma_{\rm{reac}}$ [mb]}} &

\parbox[t]{0.85cm}{\centering{ $\chi^2/F$}} \\

\hline
$^{110}$Cd & 15.6 & 1.195 & 1.046 & 362.3 & 5.495 & \multicolumn{3}{c}{$-$} &
32.1 & 1.563 & 0.344 & 71.4 & 7.617 & 456 & 0.51 \\
          &      & 1.557 & 1.000\footnote{fixed}
                                  & 411.9 & 5.251 & \multicolumn{3}{c}{$-$} &
29.8 & 1.223 & 0.682 & 83.3 & 6.473 & 506 & 0.70 \\
	  & 18.8 & 1.389 & 0.995 & 362.2 & 5.226 & \multicolumn{3}{c}{$-$} &
32.5 & 1.380 & 0.484 & 80.0 & 6.893 & 788 & 0.32 \\
$^{116}$Cd & 15.6 & 1.602 & 0.955 & 367.8 & 5.067 & \multicolumn{3}{c}{$-$} &
19.2 & 0.613 & 1.291 & 37.9 & 5.914 & 609 & 0.22 \\
           &      & 1.441 & 1.000\footnotemark[1]
                                  & 379.2 & 5.303 & \multicolumn{3}{c}{$-$} &
20.8 & 1.303 & 0.677 & 63.8 & 6.916 & 536 & 0.27 \\
           & 18.8 & 1.348 & 1.001 & 356.1 & 5.310 & \multicolumn{3}{c}{$-$} &
39.4 & 1.366 & 0.472 & 90.8 & 6.928 & 832 & 0.61 \\
\hline
$^{106}$Cd &  15.6  &   1.378 & 0.987  & 367.9 & 5.164 & -2.9 &
1.748 & 0.347 & 84.8 & 1.263 & 0.207 & 90.9 & 6.127 & 349 & 1.21 \\
          &   18.9  &   1.370 & 0.987  & 365.7 & 5.164 & -2.9 &
1.748 & 0.347 & 84.8 & 1.263 & 0.207 & 90.9 & 6.127 & 749 & 1.43 \\ 
$^{112}$Sn&          18.8 & 1.226& 1.004 & 340.6 & 5.261 & -3.1 & 1.737 &
0.341 & 89.1 & 1.252 & 0.218 & 97.2 & 6.192 & 695 & 0.77 \\
\hline
\end{tabular}
\end{table*}
\end{center}

It is well-known that there are ambiguities in the determination of the
optical potential at energies around and especially below the Coulomb
barrier. We do not consider here the so called ``family problem'' which means
that almost identical angular distributions are calculated from potentials
where the depth of the real part is increased or decreased in discrete steps
by about 30\,\%. This problem has been discussed in detail in \cite{moh97},
and its influence on $\alpha$-induced reaction cross sections for $^{106}$Cd
will be one focus of the separate study of $^{106}$Cd \cite{gal11}. We
restrict ourselves here to real potentials with $J_R \approx
350$\,MeV\,fm$^3$; these volume integrals are consistent with results which
are derived at higher energies without ambiguities \cite{atz96}.

With the above restriction for the volume integral $J_R$, the angular
distributions at 19\,MeV can be fitted satisfactorily. The reproduction
of the data is excellent, and as expected, $\chi^2/F$ values below 1.0 are
found. The width parameters remain very close to unity (deviation less than
1\,\%). The strength parameters $\lambda$ and the resulting volume integrals
$J_R$ are slightly larger by a few per cent than found for neighboring
semi-magic nuclei. The imaginary parts have volume integrals around $J_I
\approx 80 - 90$\,MeV\,fm$^3$, again somewhat larger than for neighboring
semi-magic nuclei. Such a behavior is expected from the larger absorption and
increased reaction cross section; e.g., the total reaction cross sections
$\sigma_{\rm{reac}}$ around 19\,MeV are about 10\,\% smaller for the semi-magic
even-even nuclei $^{112,124}$Sn (see \cite{moh10}) compared to non-magic
even-even $^{106,110,116}$Cd.

Unfortunately, the situation changes for the angular distributions at
16\,MeV. Here the best-fit potentials require width parameters $w$ which
deviate by about 5\,\% from unity ($w = 1.046$ for $^{110}$Cd, $w = 0.955$ for
$^{116}$Cd). However, the fit quality remains almost the same if the width
parameter $w$ is kept fix at $w = 1.0$. So it must be noted that the angular
distributions at 16\,MeV are not sufficiently sensitive to the width parameter
$w$ of the potential. Instead, the fits provide a so called ``one-point
potential'' \cite{bad78,moh97,sig00,fer10}. The smaller (larger) width
parameter $w$ is compensated by a larger (smaller) strength parameter
$\lambda$ leading to a fixed potential depth at a large radius (e.g., a value
$R_{0.2}$ where the real potential depth is 0.2\,MeV is derived in
\cite{bad78} from the analysis of elastic scattering excitation functions). We
show two calculations in Fig.~\ref{fig:ruth} using the adjusted values for $w$
and using the fixed value $w = 1.0$; the parameters of both calculations are
also listed in Table \ref{tab:local}. Although the differences between the two
calculations are small, it must be noted that the derived total reaction cross
sections $\sigma_{\rm{reac}}$ differ by about 10\,\%. Because the width
parameter $w$ is nicely determined to be close to unity from the 19\,MeV data,
we prefer the total reaction cross sections from the calculations with $w=1.0$
($\sigma_{\rm{reac}} = 506$\,mb for $^{110}$Cd and 536\,mb for $^{116}$Cd),
and we assign an uncertainty of 10\,\% for $\sigma_{\rm{reac}}$ from the
16\,MeV data. The total cross sections at 19\,MeV are well-defined with an
uncertainty of about 3\,\% (discussion of uncertainties see also
\cite{moh10}).

In addition to the above two calculations with the adjusted width parameter
$w$ and the fixed $w = 1.0$, a third calculation has been performed using the
potential which was derived at the higher energy of 19\,MeV. For both
$^{110}$Cd and $^{116}$Cd it is found that the calculated cross sections at
backward angles are slightly larger than the experimental values. This clearly
indicates that a slight energy dependence of the potential is required to
reproduce the angular distributions at both energies.

Further restrictions on the $\alpha$-nucleus potential can be derived from the
analysis of excitation functions, see Sec.~\ref{sec:exci}.

\section {Global Optical Model predictions} 
\label{sec:global}

In the present work the following open access regional and global alpha
nucleus optical potential parameterizations are considered: the recent
regional potential of Avrigeanu \textsl{et al.} \cite{avr09} and the global
potential of McFadden and Satchler \cite{mcf}. 

The regional optical potential (ROP) of Avrigeanu \textsl{et al.}
\cite{avr03} was derived starting from a semi-microscopic analysis, using the double
folding model \cite{kho94}, based on alpha-particle elastic scattering on A
$\approx$ 100 nuclei at energies below 32 MeV. The energy-dependent
phenomenological imaginary part of this semi-microscopic optical potential
takes into account also a dispersive correction to the microscopic real
potential. A small revision of this ROP and especially the use of local parameter sets were able to describe the
variation of the elastic scattering cross sections along the Sn isotopic chain \cite{avr_ad}. A further step to include all available $\alpha$-induced reaction cross sections below the Coulomb barrier has recently been carried out \cite{avr09}. First, the ROP based entirely on $\alpha$ particle elastic scattering \cite{avr03} was extended to $A\sim$ 50-120 nuclei and energies from $\sim$ 13 to 50 MeV. Secondly, an assessment of available $(\alpha,\gamma)$, $(\alpha,n)$ and $(\alpha,p)$ reaction cross sections on target nuclei ranging from $^{45}$Sc to $^{118}$Sn at incident energies below 12 MeV was carried out.
In the present study we use the potential from a review paper \cite{avr09}. A minor revision of this potential has been suggested very recently in \cite{avr10}. 

The global potential of McFadden and Satchler \cite{mcf} is fitted to the numerous alpha elastic scattering experiments done on nuclei between O and U at alpha energies of 24.7 MeV in the 60`s. Fits were obtained using a four-parameter Woods-Saxon potential. This simple potential is widely used for reaction rate calculations and for $p$ process reaction flow simulations \cite{NON_SMOKER}.

\begin{center}
\begin{table*}
\caption{Regional optical potential parameters calculated from Table 3 of \cite{avr09}.}
\setlength{\extrarowheight}{0.2cm}
\begin{tabular}{cccccccccccccc}
\hline
\multicolumn{1}{c}{}  
&
\multicolumn{1}{c}{} 
&
\multicolumn{3}{c}{Real part}  
&
\multicolumn{6}{c}{Imaginary part}  \\

\parbox[t]{1cm}{\centering{Nucleus   }} &

\parbox[t]{0.7cm}{\centering{E$_{c.m.}$ [MeV]   }} &

\parbox[t]{0.65cm}{\centering{ V$_R$ [MeV]}} &
\parbox[t]{0.65cm}{\centering{ r$_R$ [fm]}} &
\parbox[t]{0.65cm}{\centering{ a$_R$ [fm]}} &

\parbox[t]{0.65cm}{\centering{ W$_V$ [MeV]}} &
\parbox[t]{0.65cm}{\centering{ r$_V$ [fm]}} &
\parbox[t]{0.65cm}{\centering{ a$_V$ [fm]}} &
\parbox[t]{0.65cm}{\centering{ W$_S$ [MeV]}} &
\parbox[t]{0.65cm}{\centering{ r$_s$ [fm]}} &
\parbox[t]{0.65cm}{\centering{ a$_s$ [fm]}} &
\parbox[t]{0.65cm}{\centering{ $\sigma_{\rm{reac}}$ [mb]}} \\
\hline
$^{110}$Cd &15.6           & -134.2  &1.367 & 0.636 & -6.2  &1.34 & 0.50 &
21.6 & 1.52& 0.374 & 377 \\
	         &18.8           & -125.7  &1.405 & 0.602 & -9.8  &1.34 &
0.50 & 16.9 & 1.52& 0.374 & 773 \\	
$^{116}$Cd &15.6           & -134.0  &1.367 & 0.637 & -6.0  &1.34 & 0.50 &
21.9 & 1.52& 0.368 & 418 \\
                     &18.8           & -125.6  &1.406 & 0.602 & -9.6  &1.34 &
0.50 & 17.3 & 1.52& 0.368 & 821 \\
\hline
$^{106}$Cd &15.6           & -134.4  &1.367 & 0.635 & -6.4  &1.34 & 0.50 &
21.4 & 1.52& 0.379 & 348 \\     
                     &18.9           & -125.5  &1.407 & 0.600 & -10.1 &1.34 &
0.50 & 16.5 & 1.52& 0.379 & 754 \\
$^{112}$Sn &18.8           & -125.9  &1.406 & 0.601 & -9.8  & 1.34 & 0.50 &
17.0 & 1.52 & 0.372 & 706 \\
\hline
\end{tabular}
\end{table*}
\end{center}

The results of the model calculations are compared with the experimental data in Figure~\ref{fig:ruth}.  For a strict
comparison between the potentials a $\chi^2$ analysis has been done.
The resulting $\chi^2$ parameters can be found in Table \ref{tab:chi}. 

It is interesting to note that the ROP of \cite{avr09} is almost
perfect for the $^{106}$Cd case whereas it slightly overestimates the elastic
scattering cross sections at backward angles for $^{110,116}$Cd. This is also
seen in the analysis of the backward angle excitation functions (see next
Sec.~\ref{sec:exci}). It is obvious that evolution of the cross sections along
the cadmium isotopic chain cannot be reproduced exactly by the ROP under these
circumstances. Nevertheless, as discussed in detail in Sec.~\ref{sec:ratio},
the cross section ratios are very sensitive to the chosen potential, and thus
these ratios are able to provide some hints on possible improvements of the
ROP. 

\begin{center}
\begin{table*}
\caption{
\label{tab:chi}
$\chi^2_{red}$ of predictions using different global and regional
parameterizations compared with the angular distributions studied in the
present work.
} 
\setlength{\extrarowheight}{0.2cm}
\begin{tabular}{ccccccccccccr}
\hline
\parbox[t]{3cm}{\centering{Parameterization}} &
\multicolumn{2}{c}{$^{110}$Cd($\alpha,\alpha$)}  
&
\multicolumn{2}{c}{$^{116}$Cd($\alpha,\alpha$)}  
&
\multicolumn{2}{c}{$^{106}$Cd($\alpha,\alpha$)}  
&
\multicolumn{1}{c}{$^{112}$Sn($\alpha,\alpha$)} 
&
\parbox[t]{1.0cm}{\centering{Ref.}} \\

\parbox[t]{0.85cm}{\centering{   }} &
\parbox[t]{0.85cm}{\centering{ 15.6 MeV}} &
\parbox[t]{0.85cm}{\centering{ 18.9 MeV}} &

\parbox[t]{0.85cm}{\centering{ 15.6 MeV}} &
\parbox[t]{0.85cm}{\centering{ 18.8 MeV}} &

\parbox[t]{0.85cm}{\centering{ 15.6 MeV}} &
\parbox[t]{0.85cm}{\centering{ 18.8 MeV}} &
\parbox[t]{0.85cm}{\centering{ 18.8 MeV}} \\
\hline
Local            &  0.51-0.70 & 0.32 & 0.22-0.27 & 0.61& 1.21 & 1.43 & 0.77 &
this work \\
Avrigeanu         & 40.8 & 23.8 & 16.1 & 42.1 & 2.44  & 2.14& 2.25 & \cite{avr09} \\
McFaddden        &  85.8 & 62.9 & 45.8 & 103.0 & 38.8 & 54.2 & 61.3 & \cite{mcf} \\
\hline
\end{tabular}
\end{table*}
\end{center}

\section{Excitation functions at backward angles}
\label{sec:exci}
\begin{center}
\begin{figure*}
\resizebox{1.8\columnwidth}{!}{\rotatebox{0}{
\includegraphics[clip=]{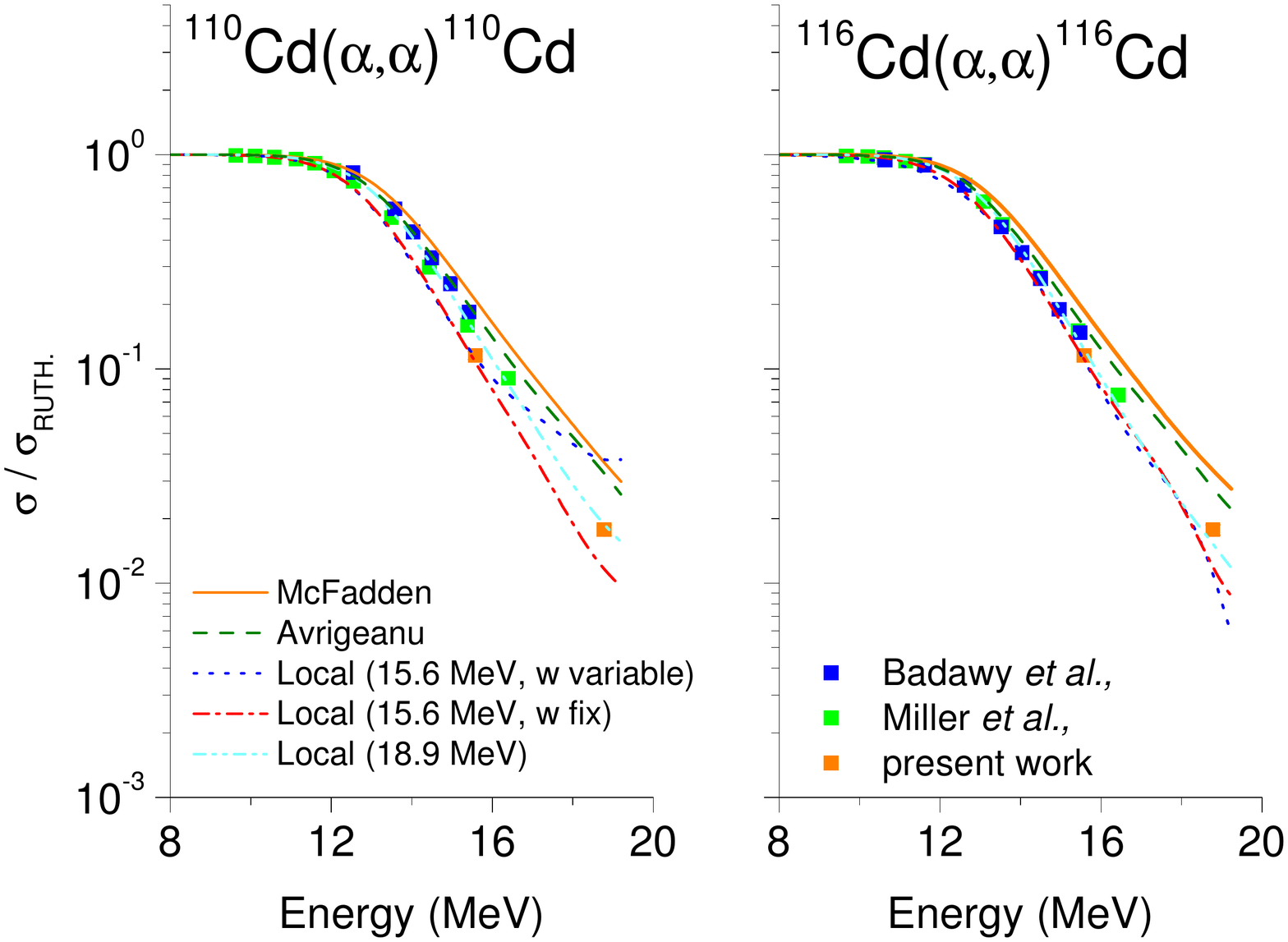}
}
}
\caption{\label{fig:exci}
(Color online.) Excitation function of $^{110}$Cd($\alpha,\alpha$)$^{110}$Cd (left) and
$^{116}$Cd($\alpha,\alpha$)$^{116}$Cd (right) reactions. Experimental data are taken
from \cite{bad78,mil81}. Further discussion see text.
}
\end{figure*}
\end{center}

Excitation functions of elastic scattering at very backward angles have been
measured by three groups \cite{eis74,bad78,mil81}. Numerical data are not
available, and all data have to be read from the given figures in
\cite{eis74,bad78,mil81}. Ref.~\cite{eis74} studies only $^{110}$Cd, and the
figure is hard to decipher. So we restrict ourselves here to the excitation
functions in Refs.~\cite{bad78,mil81}: Badawy {\it et al.}\ \cite{bad78} have
measured excitation functions at the very backward angle of
$\vartheta_{\rm{lab.}} = 178.6^\circ \approx \vartheta_{\rm{c.m.}}$ from about
10.5\,MeV to 15.5\,MeV in the c.m.\ system. Miller {\it et al.}\ \cite{mil81}
show data at $\vartheta_{\rm{lab.}} = 175^\circ$ from about 9.5\,MeV to
16.5\,MeV. Because the Rutherford-normalized cross section at very backward
angles is almost constant, we present both data sets at $175^\circ$
\cite{mil81} and $178.6^\circ$ \cite{bad78} in a common figure (the calculated
deviations because of the different angle remain below 1\,\% in the whole
energy range shown in Fig.~\ref{fig:exci}). In addition, we add the most
backward data point from our angular distributions which is also at
approximately $175^\circ$.  

As already pointed out in \cite{bad78}, it is impossible to derive an optical
potential from an excitation function at one particular angle. Instead, it is
only possible to determine an approximate strength of the imaginary potential
and a so called ``one-point'' potential for the real part. Nevertheless, the
excitation functions provide additional information, and global potentials
should be able to reproduce the measured excitation functions. In the
following we compare the predictions from the local potentials (without
further adjustment to the experimental data of the excitation function) and
from the global potentials \cite{avr09,mcf} to the experimental data
\cite{bad78,mil81}. 

For $^{110}$Cd the agreement between the experimental data of \cite{bad78} and
\cite{mil81} is not good; the data of \cite{bad78} are slightly higher than
the data of \cite{mil81}. The result from the present analysis of the full
angular distribution is a few per cent lower than \cite{mil81}. It must be
noted that the above discrepancies may -- at least partly -- be assigned to
the uncertainty of the extraction of the data from figures in
\cite{bad78,mil81}. 

The 19\,MeV local potential reproduces the excitation
function in general quite well, but slightly overestimates our lower data
point at 15.6\,MeV. The 16\,MeV local potential with the unusual width
parameter $w = 1.046$ can be excluded because of its strange energy dependence
and the strong overestimation of our 18.8\,MeV data point. The 16\,MeV local
potential with the standard width $w=1$ slightly underestimates the whole
excitation function but shows a regular energy dependence (similar to the
19\,MeV local potential). 

Both global and regional potentials also provide a regular energy
dependence; as already seen in the analysis of the angular distributions in
Fig.~\ref{fig:ruth}, both potentials overestimate the experimental data
at backward angles and thus also the excitation functions. The new potential of
\cite{avr09} is closer to the data than the old standard potential from
McFadden and Satchler \cite{mcf}. 

For $^{116}$Cd the data of \cite{bad78} and \cite{mil81} are in better
agreement. Again, our data point at 15.6\,MeV is slightly lower than the
excitation function by \cite{mil81}. The theoretical results are very similar
to the $^{110}$Cd case. Again, the 19\,MeV local potential nicely reproduces
the data. The 16\,MeV local potential with the unusual width $w = 0.955$ shows
an oscillatory energy dependence which is not visible in the experimental data, and it
underestimates the excitation function at very low energies. The 16\,MeV local
potential with $w=1$ reproduces the smooth energy dependence and gives slightly
smaller cross sections than the 19\,MeV potential.

Both global and regional potentials reproduce again the smooth energy dependence of the
data but overestimate the absolute scale. The new potential of \cite{avr09} is again closer to the data than \cite{mcf}. 

\section{Variation of the scattering cross section along isotopic and isotonic chains} 
\label{sec:ratio}

Recently, the variation of the elastic scattering cross sections along the Sn isotopic chain had been studied by Galaviz \textsl{et al.} \cite{gal05}. Complete angular distributions of the $^{112,124}$Sn($\alpha,\alpha$)$^{112,124}$Sn reactions at E$_{lab.}$ = 19.51 MeV were measured. It was found that the ratio of the elastic alpha scattering cross sections of the $^{112}$Sn and $^{124}$Sn at backward angles shows an oscillation feature. It was evident that the global alpha-nucleus potentials failed to reproduce either the amplitude and/or the phase of the oscillation pattern for backward angles \cite{gal05}. This behavior is very similar to the ratio of the Rutherford normalized cross sections of the $^{92}$Mo($\alpha,\alpha$)$^{92}$Mo and $^{89}$Y($\alpha,\alpha$)$^{89}$Y
derived by Kiss \textit{et al.}, \cite{kis09}. 

\begin{center}
\begin{figure*}
\resizebox{2.1\columnwidth}{!}{\rotatebox{0}{\includegraphics[clip=]{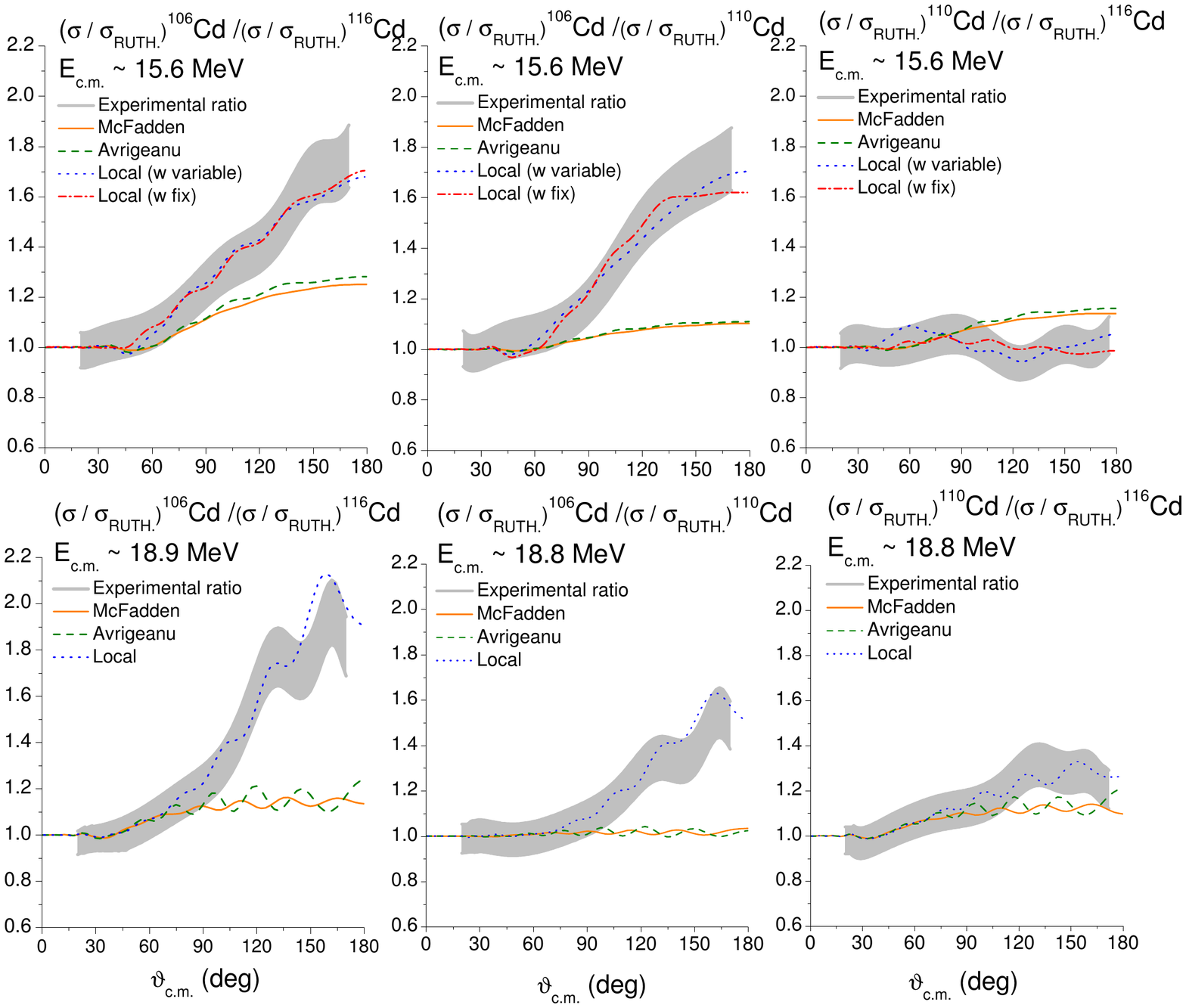}}}
\caption{
\label{fig:ratio_neutron}
(Color online.) Experimental ratio (gray area with taken into account the experimental
uncertainties) of the measured scattering cross sections at $\approx$ 15.6 MeV
($^{106}$Cd/$^{116}$Cd, $^{106}$Cd/$^{110}$Cd, $^{110}$Cd/$^{116}$Cd), 18.9
MeV ($^{106}$Cd/$^{116}$Cd) and 18.8 MeV ($^{106}$Cd/$^{110}$Cd and
($^{110}$Cd/$^{116}$Cd))versus the angle in center-of-mass frame. The cross
sections of the $^{106}$Cd($\alpha,\alpha$)$^{106}$Cd are taken from
\cite{gal_phd, kis06}. The lines correspond to the predictions using the
present local, regional \cite{avr09} and global \cite{mcf} optical potential
parameter sets. For more information see Sec. \ref{sec:local} and
\ref{sec:global}.
} 
\end{figure*}
\end{center}

In the present work, first, the behavior of the elastic alpha scattering
cross sections along the Cd isotopic ($Z$ = 48) chain is investigated at
E$_{c.m.} \approx$ 15.6 and 18.8 MeV. Although there are small differences in
the center-of-mass energies ($\leq$ 120 keV), the ratios of the Rutherford
normalized cross sections are well defined because the dominating 1/E$^2$
dependence of the scattering cross section is taken into account during Rutherford
normalization. Therefore, the ratios of Rutherford normalized cross sections
are a very sensitive test for local, regional and global alpha-nucleus
potential parameterizations. It is found that the ratio of the normalized
scattering cross sections shows an oscillation pattern at backward angles (see
Fig. \ref{fig:ratio_neutron}) similarly to the variation of the elastic
scattering cross sections along the Sn isotopic chain \cite{gal05}. The large
number of experimental points and the low uncertainties on all data sets
provide a unique probe to understand the evolution of the alpha nucleus
potential along the Cd isotopic chain.
 
\begin{center}
\begin{figure*}
\resizebox{1.6\columnwidth}{!}{\rotatebox{0}{\includegraphics[clip=]{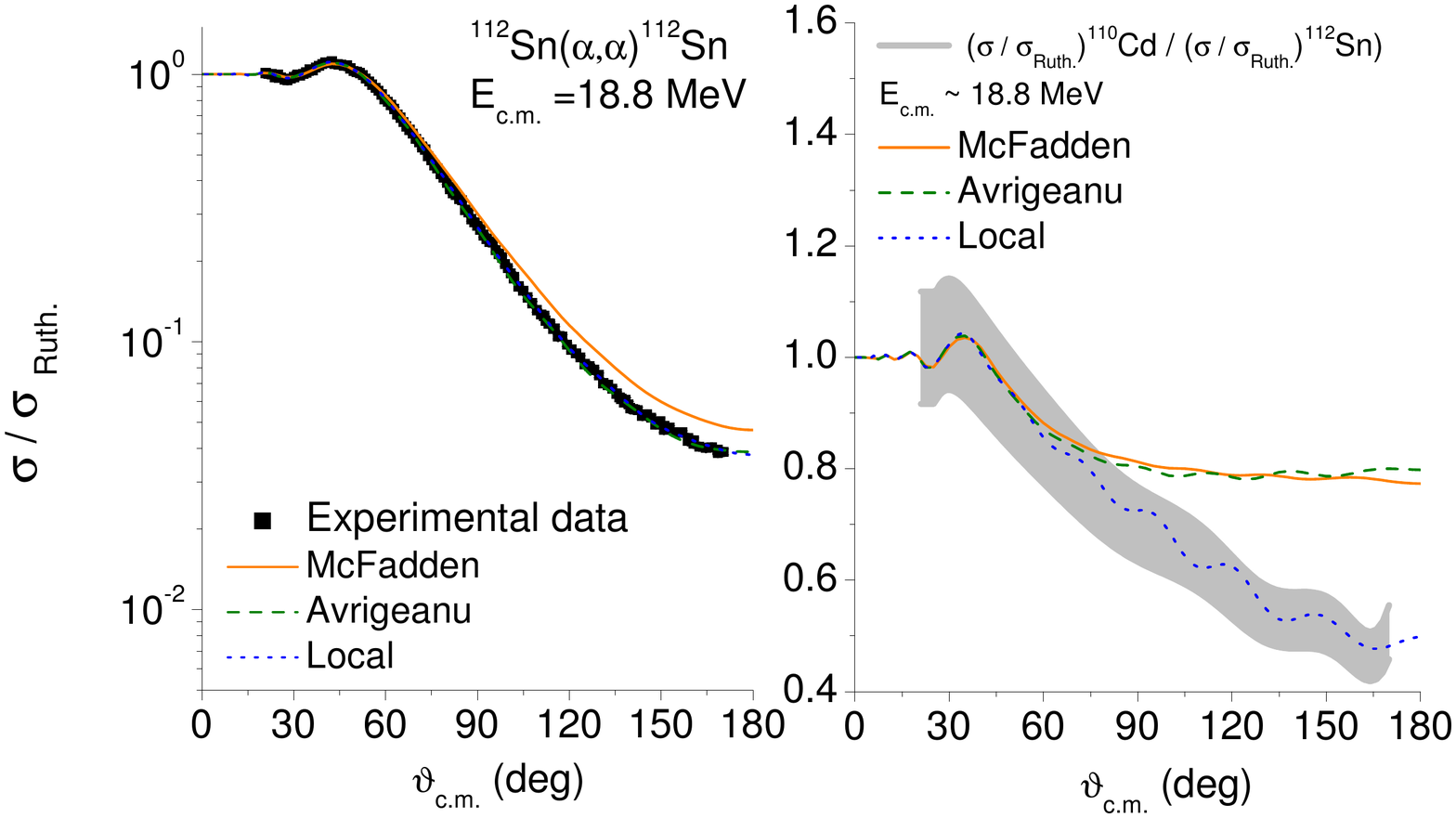}}}
\caption{
\label{fig:ratio_proton}
(Color online.) Rutherford normalized elastic scattering cross sections of
$^{112}$Sn($\alpha,\alpha$)$^{112}$Sn reaction at E$_{c.m.}$ = 18.8 MeV (left
side). Experimental ratio of the scattering cross sections
($\sigma/\sigma_{RUTH}$)$^{110}$Cd / ($\sigma/\sigma_{RUTH}$)$^{112}$Sn at
E$_{c.m.} \approx$ 18.8 MeV (gray area with taking into account the
experimental uncertainty) versus the angle in center-of-mass frame. The cross
sections of the $^{112}$Sn($\alpha,\alpha$)$^{112}$Sn are taken from
\cite{gal05}. The lines correspond to the predictions using the present local,
regional \cite{avr09} and global \cite{mcf} optical potential parameter
sets. For more information see Sec. \ref{sec:local} and \ref{sec:global}.
} 
\end{figure*}
\end{center}

Moreover,  the variation of the elastic alpha scattering cross sections along
the $N$ = 62 isotonic chain is also studied by investigating the ratio of the
$^{110}$Cd($\alpha$,$\alpha$)$^{110}$Cd and
$^{112}$Sn($\alpha$,$\alpha$)$^{112}$Sn reaction cross sections at E$_{c.m.}
\approx$ 18.8 MeV (see Fig. \ref{fig:ratio_proton}). The
$^{112}$Sn($\alpha$,$\alpha$)$^{112}$Sn is taken from \cite{gal05}. It was
found that the ratio of the elastic scattering cross sections along the $N$ =
62 isotonic chain shows a similar behavior to the one reported in
\cite{kis09}. 

In Fig. \ref{fig:ratio_neutron} and \ref{fig:ratio_proton} the experimental
ratio of the Rutherford normalized elastic scattering cross sections is
compared to the corresponding results of the regional potential of
Avrigeanu \textsl{et al.} \cite{avr09} and the global potential of
McFadden and Satchler \cite{mcf}. The grey shaded error band is a very conservative estimate which is derived
from the total errors (statistical and systematic, see discussion in Sec. II)
of the measured cross sections in Fig.~\ref{fig:ruth}. If we consider only the statistical
uncertainties, it can be clearly seen that the oscillatory patterns in the
cross section ratios are well defined by the experimental data. (Note that the
systematic uncertainty cancels out in the ratio to a large extent.) 
Figs. ~\ref{fig:ratio_neutron} and ~\ref{fig:ratio_proton} show that no regional or
global parameterization can describe correctly the amplitude and the phase of
the oscillation pattern of the experimental data at backward angles. This fact
clearly indicates that the available theoretical alpha nucleus optical
potential parameterizations have to be further improved.

A closer look at the shown ratios in Figs.~\ref{fig:ratio_neutron} and
\ref{fig:ratio_proton} and the underlying cross sections in
Fig.~\ref{fig:ruth}  provides deeper insight into the reasons for the failure
of the potentials and should lead to suggestions for improvements, in
particular for the regional potential by Avrigeanu {\it et al.}\ \cite{avr09}
with its careful parameterization of all the parameters of the potential in
dependence on the target mass number $A$, charge number $Z$, and energy
$E$. It is obvious that the mass- and energy-independent global potential by
McFadden and Satchler \cite{mcf} does a good job, but improvements within this
very limited parameter space are almost impossible. Therefore, the following
discussion focuses mainly on the ROP potential.

The ROP potential is able to reproduce the angular distributions for
$^{106}$Cd almost perfectly, and thus it also reproduces the total reaction
cross section $\sigma_{\rm{reac}}$ for $^{106}$Cd. The elastic scattering
cross sections of $^{110}$Cd and $^{116}$Cd are significantly smaller at
backward angles; this corresponds to a significantly larger
$\sigma_{\rm{reac}}$. The increase of $\sigma_{\rm{reac}}$ with neutron number
can be understood easily because of the dominance of the ($\alpha$,n) reaction
channel and its increasing cross section with increasing neutron
number. However, it is surprising that there is a strong change from
$^{106}$Cd to $^{110}$Cd and only a much smaller change from $^{110}$Cd to
$^{116}$Cd: the ratio of elastic scattering cross sections at backward angles
is about 1.5 for $^{106}$Cd/$^{110}$Cd whereas it is only about 1.25 for
$^{110}$Cd/$^{116}$Cd. Any global potential with a smooth mass dependence like
e.g.\ \cite{avr09} or missing mass dependence \cite{mcf} will fail to
reproduce the cross section ratios in Fig.~\ref{fig:ratio_neutron}; the
calculated ratios are about 1.1 in all cases and smaller than the experimental
results. The apparently different behavior of $^{106}$Cd may be understood
from a weak subshell closure of the $g_{7/2^+}$ neutron shell at $N =
58$. Although one textbook reference of the shell model \cite{kli52} shows in
its Fig.~1 that the $d_{5/2^+}$ subshell is slightly lower than the
$g_{7/2^+}$ subshell (and thus one should find a subshell closure at $N = 56$
instead of $N = 58$), there is some evidence from the ground state spins of $J
= 5/2^+$ for neighboring $N = 59$ nuclei like $^{105}$Pd, $^{107}$Cd,
$^{109}$Sn that the $g_{7/2^+}$ neutron subshell is filled at $N = 58$. Note
that the lowering of the $g_{7/2^+}$ subshell below the $d_{5/2^+}$ subshell
is well-established for the proton subshells \cite{kli52}. This weak subshell
closure may explain the relatively small total reaction cross section of
$^{106}$Cd. 

The experimental ratio of elastic scattering cross sections at backward angles
is about 0.5 between $^{110}$Cd and $^{112}$Sn. Again, this can be understood,
but the argument is different: the neutron numbers are the same for $^{110}$Cd
and $^{112}$Sn and cannot have strong impact on the ($\alpha$,n) or total
reaction cross section. However, $^{112}$Sn is a semi-magic nucleus with $Z =
50$, and thus the total reaction cross section $\sigma_{\rm{reac}}$ is smaller
compared to neighboring nuclei, and the elastic scattering cross section is
larger. Again, such a behavior cannot be reproduced by any potential with a
smooth (or even missing) mass dependence.

This leads to the following recommendations for improvements. In addition to a
smooth dependence on the mass number $A$ and charge number $Z$, a further
dependence on shell closures (e.g.\ parametrized by the distance to a closed
shell) should be included in global parameterizations of $\alpha$-nucleus
potentials. This may be complemented by a further dependence of the
neutron-to-proton ratio $N/Z$. The above recommendation may be also interpreted
as a guide to the experimentalist for further experiments on non-magic
nuclei. Note that only very few data on non-magic nuclei have entered into the
determination of the global potential \cite{avr09} above $A > 80$ (see their
Table 2); this may also explain that \cite{avr09} nicely reproduce the data
for the semi-magic $^{112}$Sn but is not able to describe the data for the
non-magic $^{110,116}$Cd with the same accuracy.

\section{Summary} 
\label{sec:summ}
In the present work angular distributions of elastically scattered alpha
particles on $^{110,116}$Cd have been measured at E$_{lab.}$ = 16.14 MeV and
19.46 MeV to provide a sensitive test for global parameterizations of the
$\alpha$-nucleus potential used in $p$ process network calculations. The
measured data cover the full angular range and have small uncertainties of
about $3-4$\,\% over the whole angular range.

A local fit to the angular distributions using a folding potential in the real
part and a surface Woods-Saxon imaginary part reproduces all measured angular
distributions with high accuracy ($\chi^2/F < 1$). The volume integrals are slightly higher
than for neighboring semi-magic nuclei. The best-fit potential at 16\,MeV
shows an unusual width parameter for $^{110}$Cd and $^{116}$Cd and does not
describe the measured excitation functions at backward angles. Very similar
fits (also with $\chi^2/F < 1$) can be obtained using the standard width $w =
1$. 

The regional and global potentials by \cite{avr09} and \cite{mcf} are able to describe the
angular distributions with relatively small deviations although both global
potentials overestimate the data at backward angles. In all cases the new
potential by \cite{avr09} is closer to the experimental data than the
potential of \cite{mcf}. The same conclusion is found for the
excitation functions at backward angles which are available from literature
\cite{bad78,mil81}. However, the situation becomes worse for the evolution of
the potentials along isotopic and isotonic chains. The measured ratio of cross
sections cannot be reproduced by any regional and global potential because the deviations
at backward angles are amplified in the ratios. A reason for this problem may
be the influence of shell closures which are not taken into account in the
parameterizations of \cite{avr09} or \cite{mcf}.

Since modeling explosive nucleosynthesis scenarios requires reaction rates on
large number of reactions involving thousands of nuclei, the $\alpha$-nucleus
potential has to be known in a wide region. The reliability of the
extrapolation to unstable nuclei has to be tested by measuring the elastic
scattering cross sections on several nuclei along isotopic and isotonic
chains. The ratio of Rutherford normalized cross sections along isotopic or
isotonic chains is a very sensitive observable for the quality of
$\alpha$-nucleus potentials, and it should be used in further work to restrict
global parameterizations of $\alpha$-nucleus potentials.

Further systematic experimental elastic alpha scattering studies at energies
around the Coulomb barrier are essential, in particular on intermediate mass
and heavy nuclei without shell closures. The experimental scattering data
should be complemented by data on $\alpha$-induced reaction cross
sections in the same energy region. Scattering and reaction data have to enter
into theoretical studies leading eventually to a robust global $\alpha$-nucleus
potential which is able to describe all observables with reasonable accuracy.

\begin{acknowledgments}
This work was supported by the EUROGENESIS research program, by OTKA (NN83261,
K068801), by the European Research Council (grant agreement no. 203175) and by the Joint Institute for Nuclear Astrophysics (NSF grant
PHY0822648). G. G. K. and D. G. acknowledge the support of the Spanish CICYT under the
project FPA2005-02379 and MEC Consolider project CSD2007-00042.  Gy. Gy\"urky
acknowledges support from the Bolyai grant. D. Galaviz is acknowledges support from Juan de la Cierva
grant (Spanish Ministry of Science). This work was also supported by the Scientific 
and Technological Research Council of Turkey (TUBITAK) Grant No: 108T508 (TBAG1001) 
and 109T585 (under the EUROGENESIS research program). Fruitful discussions with M. Avrigeanu
is greatfully acknowledged.
\end{acknowledgments}


\begin{thebibliography}{}

%
\bibitem{woo78} S. E. Woosley and W. M. Howard, Astrophys. J. Suppl.\ \textbf{36}, 285 (1978).
%
\bibitem{arn03} M. Arnould and S. Goriely, Phys.\ Rep.\ \textbf{384}, 1 (2003).
%
\bibitem{rau06} T. Rauscher, Phys. Rev. C \textbf{73}, 015804 (2006).
%
\bibitem{rap06} W. Rapp, J. G\"orres, M. Wiescher, H. Schatz and F. K\"appeler, Astrophys. J. \textbf{653}, 474 (2006).
%
\bibitem{rau10} T. Rauscher, International Journal of modern Physics E (arXiv: 1010.4283v1).
%
\bibitem{nai08} C. Nair, A. R. Junghans, M. Erhard, D. Bemmerer, R. Beyer,  E. Grosse, K. Kosev, M. Marta, G. Rusev, K. D. Schilling, R. Schwengner, and A. Wagner, Phys. Rev. C \textbf{81}, 055806 (2010).
%
\bibitem{moh08} P. Mohr, Zs. F\"ul\"op and H. Utsunomiya, Eur. Phys. J. A \textbf {32}, 357  (2007).
%
\bibitem{rau09}
T. Rauscher, G. G. Kiss, Gy. Gy\"urky, A. Simon, Zs. F\"ul\"op, E. Somorjai,
\prc {\bf 80}, 035801 (2009).
%
\bibitem{ful96} Zs. F\"ul\"op,  \'A. Z. Kiss, E. Somorjai, C. E. Rolfs, H.P. Trautvetter, T. Rauscher and H. Oberhummer,  Z. Phys.\ A {\bf 355}, 203  (1996).
%
\bibitem{rap01}  W. Rapp, M. Heil, D. Hentschel, F. K\"appeler, R. Reifarth, H. J. Brede, H. Klein and T. Rauscher,  Phys. Rev. C {\bf 66}, 015803 (2002).
%
\bibitem{gyu06} Gy.\,Gy\"urky, G. G. Kiss, Z. Elekes, Zs. F\"ul\"op, E. Somorjai, A. Palumbo, J. G\"orres, H. Y. Lee, W. Rapp, M. Wiescher, N. \"Ozkan, R. T. G\"uray, G. Efe and T. Rauscher,  Phys. Rev. C {\bf 74}, 025805  (2006).
%
\bibitem{ozk07}  N. \"Ozkan, G. Efe, R. T. G\"uray, A. Palumbo, J. G\"orres, H. Y. Lee, L. O. Lamm, W. Rapp, E. Stech, M. Wiescher, Gy. Gy\"urky, Zs. F\"ul\"op and E. Somorjai,  Phys. Rev. C {\bf 75}, 025801  (2007).
%
\bibitem{cat08} I. Cata-Danil, D. Filipescu, M. Ivascu, D. Bucurescu, N. V. Zamfir, T. Glodariu, L. Stroe, G. Cata-Danil, D. G. Ghita, C. Mihai, G. Suliman and T. Sava, Phys. Rev. C {\bf 78}, 035803 (2008).
%
\bibitem{yal09} C. Yalcin,  R. T. G\"uray, N. \"Ozkan, S. Kutlu, Gy. Gy\"urky, J. Farkas, G. G. Kiss, Zs. F\"ul\"op, A. Simon, E. Somorjai and T. Rauscher,  Phys. Rev. C {\bf 79}, 065801 (2009).
%
\bibitem{som98} E. Somorjai, Zs. F\"ul\"op, \'A. Z. Kiss, C.E. Rolfs, H.P. Trautvetter, U. Greife, M. Junker, S. Goriely, M. Arnould, M. Rayet, T. Rauscher and H. Oberhummer, Astron.\ Astrophys.\
{\bf 333} 1112 (1998).
%
\bibitem{gyu10} Gy. Gy\"urky, Z. Elekes, J. Farkas, Zs. F\"ul\"op, Z. Hal\'asz,
G. G. Kiss, E Somorjai, T Sz\"ucs, R T G\"uray, N \"Ozkan,
C Yalcin and T Rauscher, J. Phys. G. {\bf 37}, 115201 (2010).
%
\bibitem{kis10} G. G. Kiss, T. Rauscher, T. Sz\"ucs, Zs. Kert\'esz, Zs. F\"ul\"op, Gy. Gy\"urky, C. Fr\"ohlich, J. Farkas, Z. Elekes and E. Somorjai, Phys. Lett. B.  {\bf 695}, 419 (2011).
%
\bibitem{moh97} P.\ Mohr, T.\ Rauscher, H.\ Oberhummer, Z.\ M\'at\'e, Zs.\ F\"ul\"op, E.\ Somorjai, M.\ Jaeger and G.\ Staudt, 
\prc {\bf 55}, 1523 (1997).
%
\bibitem{ful01} Zs.\ F\"ul\"op, Gy.\ Gy\"urky, Z.\ M\'at\'e, E.\ Somorjai, L.\ Zolnai, D.\ Galaviz, M.\ Babilon, P.\ Mohr, A.\ Zilges, T.\ Rauscher, H. Oberhummer and G. Staudt, 
\prc {\bf 64}, 065805 (2001).
%
\bibitem{gal_phd} D. Galaviz,  Ph.D. thesis, TU Darmstadt (2004).
%
\bibitem{kis06} G.\ G.\ Kiss, Zs.\ F\"ul\"op, Gy.\ Gy\"urky, Z.\ M\'at\'e, E.\ Somorjai, D.\ Galaviz, A.\ Kretschmer, K.\ Sonnabend and A.\ Zilges,  
Eur.\ Phys.\ J.\ {\bf 27}, 197 (2006).
%
\bibitem{gal05} D.\ Galaviz, Zs.\ F\"ul\"op, Gy.\ Gy\"urky, Z.\ M\'at\'e, P.\ Mohr, T.\ Rauscher, E.\ Somorjai, and A.\ Zilges, 
\prc {\bf 71}, 065802 (2005).
%
\bibitem{pal08} A.Palumbo, W.Tan, J.Görres, M.Wiescher, Zs. F\"ul\"op, Gy. Gy\"urky, G. G. Kiss, E. Somorjai, D. Galaviz, N.\"Ozkan, R.T. G\"uray,
POS (NIC X) 046 (2008). 
%
\bibitem{kis09} G.\ G.\ Kiss, P. Mohr, Zs.\ F\"ul\"op, D. Galaviz, Gy.\ Gy\"urky, Z. Elekes, E. Somorjai, A. Kretschmer, K. Sonnabend, A. Zilges, and M. Avrigeanu, 
Phys. Rev. C {\bf 80}, 045807 (2009).
%
\bibitem{eis74}
Y. Eisen, E. Abramson, G. Engler, M. Samuel, U. Smilansky, Z. Vager,
Nucl. Phys. {\bf A236}, 327 (1974).
%
\bibitem{bad78}
I. Badawy, B. Berthier, P. Charles, M. Dost, B. Fernandez, J. Gastebois,
S. M. Lee,
\prc {\bf 17}, 978 (1978).
%
\bibitem{mil81}
M. Miller, A. M. Kleinfeld, A. Bockisch, K. Bharuth-Ram,
Z. Phys. A {\bf 300}, 97 (1981).
%
\bibitem{gal11}
D. Galaviz {\it et al.},
to be published.
%
\bibitem{mat89} Z.~M\'at\'e, S.~Szil\'agyi, L.~Zolnai, \AA.~Bredbacka, M.~Brenner, K.-M.~K\"allmann, and P.~Manng{\aa}rd, Acta Phys.~Hung.~{\bf 65}, 287 (1989).
%
\bibitem{kis08} G. G. Kiss, D. Galaviz, Gy. Gy\"urky, Z. Elekes, Zs.\ F\"ul\"op, E. Somorjai, K. Sonnabend, A. Zilges, P. Mohr, J. G\"orres,
M. Wiescher, N. \"Ozkan, T. G\"uray, C. Yalcin and M. Avrigeanu, AIP conf. proc. {\bf 1016}, 221 (2008).
%
\bibitem{nndc106} D. De. Frenne and A. Negret, Nuclear Data Sheets {\bf 109}, 943 (2008). 
%
\bibitem{nndc110} D. De Frenne and A. Jacobs, Nuclear Data Sheets {\bf 89}, 481 (2000).
%
\bibitem{nndc116} J. Blachot, Nuclear Data Sheets {\bf 92}, 455 (2001).
%
\bibitem{nndc112} D. De Frenne and E. Jacobs, Nuclear Data Sheets {\bf 79}, 639 (1996).
%
\bibitem {kob84} 
  A.~M.~Kobos, B.~A.~Brown, R.~Lindsay, and G.~R.~Satchler,
  Nucl.~Phys.~{\bf A425}, 205 (1984).
%
\bibitem {sat79} 
  G.~R.~Satchler and W.~G.~Love,
  Phys.~Rep.~{\bf 55}, 183 (1979).
%
\bibitem{abe93}
H. Abele and G. Staudt,
\prc {47}, 742 (1993).
%
\bibitem{atz96} 
U.\ Atzrott, P.\ Mohr, H.\ Abele, C.\ Hillenmayer, G.\ Staudt,
\prc {\bf 53}, 1336 (1996).
%
\bibitem{vri87} H.~de Vries, C.~W.~de Jager, and C.~de Vries,
  Atomic Data and Nuclear Data Tables {\bf 36}, 495 (1987).
%
\bibitem{moh10}
P.\ Mohr, D.\ Galaviz, Zs.\ F\"ul\"op, Gy. Gy\"urky, 
G.\ G.\ Kiss, E.\ Somorjai,
\prc {\bf 82}, 047601 (2010).
%
\bibitem{sig00}
C.\ Signorini, A.\ Andrighetto, M.\ Ruan, J.\ Y.\ Guo, L.\ Stroe, F.\ Soramel,
K.\ E.\ G.\ L{\"o}bner, L.\ M{\"u}ller, D.\ Pierroutsakou, M.\ Romoli,
K.\ Rudolph, I.\ J.\ Thompson, M.\ Trotta, A.\ Vitturi, R.\ Gernh{\"a}user,
A.\ Kastenm{\"u}ller,
\prc {\bf 61}, 061603(R) (2000).
%
\bibitem{fer10}
J.\ P.\ Fern{\'a}ndez-Garc{\'i}a, M.\ Rodr{\'i}guez-Gallardo,
M.\ A.\ G.\ Alvarez, A.\ M.\ Moro,
Nucl.\ Phys.\ {\bf A840}, 19 (2010).
%
\bibitem{avr09} M. Avrigeanu, A.C. Obreja, F.L. Roman, V. Avrigeanu, W. von Oertzen, At. Data Nucl. Data Tables {\bf 95}, 501 (2009).
%
\bibitem{avr10} M. Avrigeanu and V. Avrigeanu, \prc {\bf 82}, 014606 (2010).
%
\bibitem{mcf} L. McFadden and G. R. Satchler, Nucl. Phys. {\bf84}, 177 (1966).
%
%
%
%
%
%
\bibitem{kli52}
P.\ A.\ Klinkenberg,
Rev.\ Mod.\ Phys.\ {\bf 24}, 63 (1952).
%
\bibitem{NON_SMOKER} T. Rauscher, code NON-SMOKERWEB, http://nucastro.org/websmoker.html 
%
\bibitem{avr03} M. Avrigeanu, W. von Oertzen, A.J.M. Plompen and V. Avrigeanu,
Nucl. Phys. A{\bf723}, 104 (2003).
%
M. Avrigeanu, W. von Oertzen and V. Avrigeanu,
Nucl. Phys. A{\bf764}, 246 (2006).
%
\bibitem{kho94} D. T. Khoa, W. von Oertzen and H. G. Bohlen, \prc {\bf 49}, 1652 (1994).
%
\bibitem{avr_ad} M. Avrigeanu and V. Avrigeanu, Phys. Rev. C {\bf 73}, 038801 (2006).
%
\end{thebibliography}
\end{document}